\documentclass[11pt]{article}
\usepackage{amsmath,amssymb,amsfonts,bbm}

\newcommand{\be}{\begin{equation}}
\newcommand{\ee}{\end{equation}}
\newcommand{\bea}{\begin{eqnarray}}
\newcommand{\eea}{\end{eqnarray}}
\newcommand{\ba}{\begin{array}}
\newcommand{\ea}{\end{array}}

\newcommand{\N}{\mathcal{N}}

\long\def\symbolfootnote[#1]#2{\begingroup%
\def\thefootnote{\fnsymbol{footnote}}\footnote[#1]{#2}\endgroup}

\begin{document}

\thispagestyle{empty}\vspace{40pt}

\hfill{}

\vspace{128pt}

\begin{center}
    \textbf{\Large The five dimensional universal hypermultiplet and \\ the cosmological constant problem}\\
    \vspace{40pt}

    Charles A. Canestaro\symbolfootnote[1]{\tt charles.canestaro@cortland.edu, charles.canestaro@gmail.com} and Moataz H. Emam\symbolfootnote[2]{\tt moataz.emam@cortland.edu}

    \vspace{12pt}   \textit{Department of Physics}\\
                    \textit{SUNY College at Cortland}\\
                    \textit{Cortland, NY 13045, USA}\\
\end{center}

\vspace{40pt}

\begin{abstract}
We model the universe as a 3-brane embedded in five dimensional spacetime with $\N=2$ supersymmetry. The presence of the scalar fields of the universal hypermultiplet in the bulk results in a positive pressure effectively reducing the value of the cosmological constant and thereby providing a possible answer as to why the measured value of the cosmological constant is many orders of magnitude smaller than predicted from the vacuum energy. The solution allows for any number of parallel branes to exist and relates their cosmological constants (as well as matter densities and radiation pressures) to the value of the dilaton in the extra dimension. The results we find can be thought of as first order approximations, satisfying supersymmetry breaking and the Bogomol'nyi-Prasad-Sommerfield (BPS) conditions in the bulk only.
\end{abstract}

\newpage



\vspace{15pt}

\pagebreak

\section{Introduction}

The 1998 discovery that the universe's expansion is accelerating \cite{astro-ph/9805201, astro-ph/9812133} took the scientific community by storm. The term `dark energy' was coined to account for the source of the mysterious negative pressure causing the acceleration. From the point of view of the Einstein gravitational field equations, this energy is modeled by the presence of a positive cosmological constant $\Lambda$. Although many different models explaining the origin of this dark energy exist, it is somewhat accepted today that the main contributor is almost certainly the energy of the vacuum as calculated from the standard model (see, for example, \cite{1203.0307} and references within). There is a problem with this, however, as the vacuum's contribution seems to be many orders of magnitude \emph{larger} than is actually observed, at least at the present epoch. The most recent estimates (e.g. \cite{1001.4538}) put the difference to about 120 orders of magnitude; a catastrophic discrepancy that cannot simply be approximated away! Many solutions, as well as partial solutions, to this problem have been proposed. Of particular interest to us are models that embed our universe in a higher dimensional space, usually as a `string-theoretic' 3-brane. Such models of `brane cosmology' exist in abundance and range over various categories from non-supersymmetric solutions, to solutions that explain the acceleration with zero cosmological constant, and many others (see, for example, \cite{astro-ph/0208223, 1006.1850, hep-th/0009167, hep-th/0006215, hep-ph/9910498, hep-th/9910219, hep-th/0008192, 0804.2248, hep-th/0611074, 0709.3136, hep-th/0308125}).

In this paper we consider five dimensional $\N=2$ supergravity with the scalar fields of the universal hypermultiplet (UH). Our universe is modeled as a 3-brane satisfying the conditions of homogeneous and isotropic expansion, \emph{i.e.} the standard Robertson-Walker metric construction. The current work generalizes a static 3-brane solution that was found in \cite{1108.3391} to one with time dependence. We assume the simple form of constant matter density and radiation pressure in the universe, as well as a possible cosmological constant in the bulk. As such our model does not chart the entire history of the universe, but rather just an effective snapshot of a specific epoch. We find that a generalization of our result to one with multiple brane solutions (parallel universes) is quite straightforward and is in fact almost demanded by the equations. The observed numerical values of the cosmological constant, matter and radiation pressure densities are tied in to the values of the UH fields in the bulk and as such change from brane to brane. Our universe's observed values are the way they are by virtue of our presence in this particular brane rather than another. Furthermore, we find that the values of these constants are controlled by a free parameter $B$ that is not determined by the model. But a lower bound may simply be placed on it by requiring that the fields vanish at bulk infinity. The entire result can be thought of as a first order approximation in the following sense: The brane's matter and radiation contents are not included in the theory's action and as such do not couple to the gravitini; so the supersymmetry variation equations are valid only in the bulk. This opens up a variety of questions and directions of possible future research, as will be subsequently discussed.

The paper is organized as follows: In section (\ref{N=2SUGRA}) we review the five dimensional supergravity theory formulated in the language of split complex numbers. In section (\ref{StudyingSTbackground}) we introduce our metric ansatz and calculate the components of the Einstein field equations. It is shown that two possible solutions exist. Only one is discussed in this paper, the other is deferred to future work. The UH fields are found in section (\ref{TheFields}). Finally, the modified Friedmann equations are derived and the full solution is summarized in section (\ref{FriedmannFinal}).

\section{Five dimensional $\N=2$ supergravity}\label{N=2SUGRA}

The dimensional reduction of $D=11$ supergravity theory (see \cite{Emam:2010kt} for a review) over a rigid Calabi-Yau 3-fold (constant K\"{a}hler and complex structure moduli) yields an ungauged $\N=2$ supersymmetric gravity theory in $D=5$ with a matter sector comprised of four scalar fields and their superpartners; collectively known as the universal hypermultiplet. These are: the dilaton $\sigma$ (volume modulus of the Calabi-Yau space), the universal axion $\varphi$, the pseudo-scalar axion $\chi$ and its complex conjugate $\bar \chi$ \cite{Ferrara:1989ik, Cecotti:1988qn}. In \cite{1108.3391}, it was argued that another way to represent the theory is by employing split-complex numbers, as opposed to the more traditional complex representation. To do so, the axions are defined as follows
\bea
    \chi  &=& \chi _1  + j\chi _2  \nonumber\\
    \bar \chi  &=& \chi _1  - j\chi _2,
\eea
where $\left(\chi _1, \chi _2\right)$ are real functions and the `imaginary' number $j$ is defined by $j^2=+1$ but is \emph{not} equal to $\pm1$. In this representation, the bosonic action of the theory is:
\be
    S_5  = \int\limits_5 {\left[ {R\star \mathbf{1} - \frac{1}{2}d\sigma  \wedge \star d\sigma  - e^\sigma  d\chi  \wedge \star d\bar \chi  - \frac{1}{2} e^{2\sigma } \left( {d\varphi + \frac{j}{2}f} \right) \wedge \star \left( {d\varphi - \frac{j}{2}\bar f} \right)} \right]},\label{ActionSPLIT}
\ee
where $\star$ is the $D=5$ Hodge duality operator and we have defined
\bea
    f &=& \left( {\chi d\bar \chi  - \bar \chi d\chi } \right) \nonumber\\
    \bar f &=&  - f,\label{f_definition}
\eea
for brevity\footnote{It is noted that the action in this form suffers from the presence of high energy ghost-like terms. However, as we will see, these are exactly canceled in our solution and as such have no physical effect on our conclusions.}. The variation of the action yields the following field equations for $\sigma$, $\left(\chi, \bar\chi\right)$ and $\varphi$ respectively
\bea
    \left( {\Delta \sigma } \right)\star \mathbf{1} - e^\sigma  d\chi  \wedge \star d\bar \chi  - e^{2\sigma } \left( {d\varphi + \frac{j}{2}f} \right) \wedge \star\left( { d\varphi - \frac{j}{2}\bar f} \right) &=& 0 \label{Dilaton eomSPLIT}\\
    d^{\dagger} \left[ {e^\sigma  d\chi  + je^{2\sigma } \chi \left( {d\varphi + \frac{j}{2}f} \right)} \right] &=& 0 \label{Chi eomSPLIT}\\
    d^{\dagger} \left[ {e^\sigma  d\bar \chi  - je^{2\sigma } \bar \chi \left( {d\varphi - \frac{j}{2}\bar f} \right)} \right] &=& 0
    \label{Chi bar eomSPLIT}\\
    d^{\dagger} \left[ {e^{2\sigma } \left( {d\varphi + \frac{j}{2}f} \right)} \right] &=& 0\label{a eomSPLIT},
\eea
where $d^\dag$ is the adjoint exterior derivative and $\Delta$ is the Laplace-De Rahm operator. The full action is invariant under the following set of supersymmetry (SUSY) transformations of the gravitini $\psi$ and hyperini $\xi$ fermionic fields respectively $\left( {M = 0, \cdots ,4} \right)$:
\bea
    \delta _\epsilon  \psi ^1  &=& D\epsilon _1  + \frac{j}{4}e^{  \sigma } \left( {d\varphi + \frac{j}{2}f} \right)\epsilon _1  - \frac{j}{4} {{e^{\frac{\sigma }{2}} }}{{ }}d\chi \epsilon _2  \label{1SPLIT}\\
    \delta _\epsilon  \psi ^2  &=& D\epsilon _2  - \frac{j}{4}e^{  \sigma } \left( {d\varphi - \frac{j}{2}\bar f} \right)\epsilon _2  + \frac{j}{4}{{e^{\frac{\sigma }{2}} }}{{ }}d\bar \chi \epsilon _1 \label{Gravitini variationsSPLIT}\\
    \delta _\epsilon  \xi _1  &=& \frac{1}{2}\left[ {\left( {\partial _M \sigma } \right) - je^{  \sigma } \left( {\partial _M \varphi + \frac{j}{2}f_M } \right)} \right]\Gamma ^M \epsilon _1  + j\frac{{e^{\frac{\sigma }{2}} }}{{\sqrt 2 }}\left( {\partial _M \chi } \right)\Gamma ^M \epsilon _2  \label{2SPLIT}\\
    \delta _\epsilon  \xi _2  &=& \frac{1}{2}\left[ {\left( {\partial _M \sigma } \right) + je^{  \sigma } \left( {\partial _M \varphi - \frac{j}{2}\bar f_M } \right)} \right]\Gamma ^M \epsilon _2  - j\frac{{e^{\frac{\sigma }{2}} }}{{\sqrt 2 }}\left( {\partial _M \bar \chi } \right)\Gamma ^M \epsilon _1,
    \label{Hyperini variationsSPLIT}
\eea
where
\be
    D =  {d  + \frac{1}{4}\omega ^{\hat M\hat N} \Gamma _{\hat M\hat N} },\label{Covariant derivative}
\ee
is the usual covariant derivative, the $\Gamma$'s are the $D=5$ Dirac matrices, $\left(\epsilon_1,\epsilon_2\right)$ are the $\N=2$ SUSY spinors, $\omega$ is the spin connection and the hatted indices are frame indices in a flat tangent space.

\section{Spacetime background}\label{StudyingSTbackground}

From the point of view of $D=5$ SUGRA, our universe may be modeled by a 3-brane in a five dimensional bulk. This implies the embedding of the Robertson-Walker metric in five dimensional space as follows
\be
    ds_5^2  = e^{2C\sigma \left( y \right)} \left[ { - dt^2  + a^2 \left( t \right)\left( {\frac{{dr^2 }}{{1 - kr^2 }} + r^2 d\Omega ^2_2 } \right)} \right] + e^{2B\sigma \left( y \right)} b^2 \left( t \right)dy^2
\ee
where $d\Omega ^2_2 = d\theta ^2  + \sin ^2 \theta d\phi ^2$ is the line element of the unit sphere $S^2$, the quantities $C$ and $B$ are constants to be determined, $a \left( t \right)$ is the usual scale factor and $b \left( t \right)$ is a bulk scale factor. The solution we seek should have $a\left( t \right) \sim e^{Ht} $, where the positive constant $H$ is the current value of the Hubble parameter, to account for the accelerating phase of the universe. Also, since current data \cite{0901.3354} seems to imply that on a large scale our universe is essentially flat, we will then take the `curvature' factor $k$ to be zero.

The matter content of this five dimensional space is comprised of the UH fields in the bulk, represented by the following stress tensor $\left( {\mu ,\nu  = t,r,\theta,\phi} \right)$:
\bea
 T^{{\rm bulk}}_{\mu \nu}  &=&  \frac{1}{4}g _{\mu \nu} \left( {\partial _y  \sigma } \right)\left( {\partial ^y  \sigma } \right)  +\frac{1}{2}g _{\mu \nu} e^\sigma  \left( {\partial _y  \chi } \right)\left( {\partial ^y  \bar \chi } \right)
 +  \frac{1}{4}g _{\mu \nu} e^{2\sigma } \left( {\partial _y \varphi + \frac{j}{2}f_y } \right)\left( {\partial ^y \varphi - \frac{j}{2}\bar f^y } \right)  \nonumber\\
 T^{{\rm bulk}}_{yy}  &=&  \frac{1}{4}g _{yy } \left( {\partial _y  \sigma } \right)\left( {\partial ^y  \sigma } \right) -\frac{1}{2}\left( {\partial _y  \sigma } \right)\left( {\partial _y  \sigma } \right)
  +    \frac{1}{2}e^{\sigma } g _{y y }   \left( {\partial _y  \chi } \right)\left( {\partial ^y  \bar \chi } \right)-e^\sigma  \left( {\partial _y  \chi } \right)\left( {\partial _y  \bar \chi } \right) \nonumber\\
  &+&   \frac{1}{4}e^{2\sigma } g _{y y } \left( {\partial _y \varphi + \frac{j}{2}f_y } \right)\left( {\partial ^y \varphi - \frac{j}{2}\bar f^y } \right) - \frac{1}{2}e^{2\sigma } \left( {\partial _y \varphi + \frac{j}{2}f_y } \right)\left( {\partial _y \varphi - \frac{j}{2}\bar f_y } \right),\label{Bulk Stress tensor}
\eea
in addition to the usual perfect fluid stress tensor on the brane:
\be
    {\rm T}_{\mu \nu }^{{\rm 3brane}}  = \rho U_\mu  U_\nu   + P\left( {g_{\mu \nu }  + U_\mu  U_\nu  } \right),\label{PerfectFluidT}
\ee
where $\rho$ is the average mass density of matter in the 3-brane, $P$ is the thermal radiation pressure (both assumed constant in this paper), and $U$ is the four-velocity of timelike observers in the usual manner. It is unclear in the literature how an action can be constructed that will give (\ref{PerfectFluidT})\footnote{See \cite{gr-gc/9304026} and references within for various attempts to do so}, it follows that it is also unclear how to couple the parameters $\rho$ and $P$ to the gravitini. This will result, as we will see, into approximate SUSY variation equations and BPS conditions.

The independent components of the left hand side of Einstein's equation, including a brane cosmological constant $\Lambda$, as well as a possible bulk cosmological constant ${\tilde \Lambda }$, are then:
\bea
    G_{tt}  + \Lambda g_{tt}  &=& 3\left[ {\left( {\frac{{\dot a}}{a}} \right)^2  + \left( {\frac{{\dot a}}{a}} \right)\left( {\frac{{\dot b}}{b}} \right)} \right] - \frac{{3C}}{{b^2 }}e^{2\left( {C - B} \right)\sigma } F\left( \sigma  \right) - \Lambda e^{2C\sigma }  \nonumber\\
    G_{rr}  + \Lambda g_{rr}  &=&  - \left( {2\ddot aa + \dot a^2 } \right) - \frac{a}{b}\left( {\ddot ba + 2\dot a\dot b} \right) + 3C\left( {\frac{a}{b}} \right)^2 e^{2\left( {C - B} \right)\sigma } F\left( \sigma  \right) + \Lambda a^2 e^{2C\sigma }  \nonumber\\
    G_{yy}  + \tilde \Lambda g_{yy}  &=& 6C^2 \sigma '^2  - 3e^{2\left( {B - C} \right)\sigma } \left( {\frac{b}{a}} \right)^2 \left( {\ddot aa + \dot a^2 } \right) + \tilde \Lambda b^2 e^{2B\sigma }  \nonumber\\
    G_{yt}  &=& 3C\sigma '\left( {\frac{{\dot b}}{b}} \right)\,\,\,\,\,\,\,\,\,\,\,\,\,\,{\rm where}\,\,\,\,\,\,\,\,\,\,\,\,\,\,F\left( \sigma  \right) = \sigma '' + \left( {2C - B} \right)\sigma '^2,
\eea
where a prime is a derivative with respect to $y$ and a dot is a derivative with respect to time.

Generally speaking, if one requires the brane to satisfy the Bogomol'nyi-Prasad-Sommerfield condition, breaking half of the supersymmetries of the theory, then one must also necessarily require the vanishing of the variation of gravitini and hyperini backgrounds, \textit{i.e.} $\delta \psi=0$ and $\delta \xi=0$. The spin connections are:
\bea
    \omega ^{\hat t\hat r}  &=&  \dot a dr, \quad \quad\omega ^{\hat t\hat \theta }  = r\omega ^{\hat t\hat r}, \quad \quad\omega ^{\hat t\hat \phi }  = r\sin \theta \omega ^{\hat t\hat r}   \nonumber\\
    \omega ^{\hat t\hat y}  &=&  \frac{C}{b}\sigma 'e^{\left( {C - B} \right)\sigma } dt + \dot be^{\left( {B - C} \right)\sigma } dy \nonumber\\
    \omega ^{\hat r\hat y}  &=&  C\frac{a}{b}e^{\left( {C - B} \right)\sigma } \sigma 'dr, \quad \quad\omega ^{\hat \theta \hat y}  = r\omega ^{\hat r\hat y}, \quad \quad\omega ^{\hat \phi \hat y}  = r\sin \theta \omega ^{\hat r\hat y}  \nonumber\\
    \omega ^{\hat \theta \hat r}  &=& d\theta  , \quad\quad
    \omega ^{\hat \phi \hat r}  = \sin \theta d\phi  , \quad\quad
    \omega ^{\hat \phi \hat \theta }  = \cos \theta d\phi.
\eea

\section{The hyper-scalar fields in the bulk}\label{TheFields}

We begin by considering equation (\ref{a eomSPLIT}). It can be integrated once to yield
\be
    e^{2\sigma } \left( {d\varphi + \frac{j}{2}f} \right) = n dh,\label{BASIC ANSATZ}
\ee
where $h\left(y\right)$ is a harmonic function; $\Delta h = d^\dagger dh = 0$ and $n \in \mathbb{R}$. Since we require that $\delta \xi =0$, then one can write equations (\ref{2SPLIT}, \ref{Hyperini variationsSPLIT}) as follows
\be
    \left[ {\begin{array}{*{20}c}
   {\frac{1}{2}\left[ {\left( {\partial _y \sigma } \right) - je^{  \sigma } \left( {\partial _y \varphi + \frac{j}{2}f_y } \right)} \right]\Gamma ^y} & {} & {j\frac{{e^{\frac{\sigma }{2}} }}{{\sqrt 2 }}\left( {\partial _y \chi } \right)\Gamma ^y}  \\
   {} & {} & {}  \\
   {-j\frac{{e^{\frac{\sigma }{2}} }}{{\sqrt 2 }}\left( {\partial _y \bar \chi } \right)\Gamma ^y} & {} & {\frac{1}{2}\left[ {\left( {\partial _y \sigma } \right) + je^{  \sigma } \left( {\partial _y \varphi - \frac{j}{2}\bar f_y } \right)} \right]\Gamma ^y}  \\
\end{array}} \right]\left( {\begin{array}{*{20}c}
   {\epsilon_1}  \\
   {}  \\
   {\epsilon_2}  \\
\end{array}} \right) = 0,
\ee
satisfied if the determinant of the given matrix vanishes:
\be
    d\sigma  \wedge \star d\sigma  -  e^{2\sigma } \left( {d\varphi + \frac{j}{2}f} \right) \wedge \star\left( {d\varphi - \frac{j}{2}\bar f} \right) + 2e^\sigma  d\chi  \wedge \star d\bar \chi  = 0.\label{BPS condition}
\ee

Using (\ref{BASIC ANSATZ}) and (\ref{BPS condition}) into the dilaton field equation (\ref{Dilaton eomSPLIT}) gives
\be
    \left( {\Delta \sigma } \right)\star \mathbf{1} + \frac{1}{2}d\sigma  \wedge \star d\sigma  =  { \frac{{3 }}{2}}n ^2  e^{ - 2\sigma } d h \wedge \star d h.\label{Dilaton a BPS}
\ee

The axion equations (\ref{Chi eomSPLIT}, \ref{Chi bar eomSPLIT}) may also be integrated once to give
\bea
    d\chi  + nj\chi e^{ - \sigma } d h = 0 \nonumber\\
    d\bar \chi  - nj\bar \chi e^{ - \sigma } d h = 0,\label{TheCHIs}
\eea
where without loss of generality, an arbitrary constant of integration has been set to zero. The form of equation (\ref{Dilaton a BPS}) implies a solution of the form
\be
    \sigma  = m\ln h,
\ee
where $m \in \mathbb{R}$. Using this in equation (\ref{Dilaton a BPS}) immediately gives $m=n^2=1$. Integrating (\ref{TheCHIs}) leads to:
\bea
    \chi  &=& Ah^{ - nj}  \nonumber\\
    \bar \chi  &=& \bar Ah^{nj},
\eea
where $A$ is a split-complex constant of integration having the property of being null, \emph{i.e.} $\left| A \right|^2  = A\bar A = 0$, while $A$ itself is non-vanishing. This makes $\chi$ null as well. This is a property unique to the split-complex numbers and is obviously absent from the ordinary complex numbers. The reason for it is that the magnitude of a split-complex number is given by the hyperbolic modulus $\left| \chi  \right|^2  = \mathfrak{Re}^2\left(\chi\right)  - \mathfrak{Im}^2 \left(\chi\right) $. So $\chi$ becomes null simply if $\mathfrak{Re}\left(\chi\right)=\pm\mathfrak{Im}\left(\chi\right)$.
Based on these results, one finds that the quantity $f$ defined in (\ref{f_definition}) vanishes since it is proportional to $\left| A \right|^2$, and we are left with an easily found form for the universal axion as follows:
\be
    \varphi =  - \frac{n}{h} + {\rm constant}.
\ee

Finally, solving the harmonic condition on $h$ leads to:
\be
    h\left( y \right) = \left( {Qy + 1} \right)^{\frac{1}{{4C - B + 1}}}.\label{Harmonic}
\ee
Clearly all solutions depend on the magnitude of the constant $Q$. We will see that a non-vanishing $Q$ is exactly what's needed to reduce the expected value of the brane's cosmological constant. Based on these results, the independent components of the stress tensor (including both the brane and bulk contributions) become
\bea
    T_{tt}  &=&  - \frac{1}{{2b^2 }}e^{2\left( {C - B} \right)\sigma } \sigma '^2  + \rho e^{2C\sigma }  \nonumber\\
    T_{rr}  &=& \frac{1}{2}\left( {\frac{a}{b}} \right)^2 e^{2\left( {C - B} \right)\sigma } \sigma '^2  + Pa^2 e^{2C\sigma }  \nonumber\\
    T_{yy}  &=&  - \frac{1}{2}\sigma '^2  \nonumber\\
    T_{yt}  &=& 0.
\eea

The Einstein equation $G_{yt}=T_{yt}$ immediately implies that either $C$ or $\dot b$ must vanish (or both). As such there seem to be two possible solutions. In what follows, we will focus only on the $\dot b=0$ scenario and defer study of the $C=0$ case to future work.

As mentioned earlier, the fact that ${\rm T}_{\mu \nu }^{{\rm 3brane}}$ is put in the model ``by hand'', rather than naturally arising from the theory's action, leads to the problem that the brane matter and radiation densities do not couple to the supersymmetry fermions. The hyperini variation equations feature the bulk content as follows:
\bea
    \delta _y  \xi ^1  &=& \frac{1}{2}\left( {1 - nj} \right)\left( {\partial _y \sigma } \right)\Gamma ^y \epsilon _1  - \frac{n}{{\sqrt 2 }}Ae^{\left( {\frac{1}{2} - nj} \right)\sigma } \left( {\partial _y \sigma } \right)\Gamma ^y \epsilon _2  \nonumber\\
    \delta _y  \xi ^2  &=& \frac{1}{2}\left( {1 + nj} \right)\left( {\partial _y \sigma } \right)\Gamma ^y \epsilon _2  - \frac{n}{{\sqrt 2 }}\bar Ae^{\left( {\frac{1}{2} + nj} \right)\sigma } \left( {\partial _y \sigma } \right)\Gamma ^y \epsilon _1.
\eea

Requiring the vanishing of these leads to $A = \left( {1 + nj} \right)$, $\epsilon _1  = A\epsilon _2 $. On the other hand, the gravitini variations:
\bea
    \delta _t \psi ^s  &=& \left( {\partial _t \epsilon _s } \right) + \frac{1}{2}\frac{C}{b}\sigma 'e^{\left( {C - B} \right)\sigma } \Gamma _{\hat t\hat y} \epsilon _s, \quad\quad {\rm where} \quad s=1,2  \nonumber\\
    \delta _r \psi ^s  &=& \left( {\partial _r \epsilon _s } \right) + \frac{1}{2}\dot a\Gamma _{\hat t\hat r} \epsilon _s  + \frac{C}{2}\frac{a}{b}\sigma 'e^{\left( {C - B} \right)\sigma } \Gamma _{\hat r\hat y} \epsilon _s  \nonumber\\
    \delta _\theta  \psi ^s  &=& \left( {\partial _\theta  \epsilon _s } \right) + \frac{1}{2}r\dot a\Gamma _{\hat t\hat \theta } \epsilon _s  - \frac{1}{2}\Gamma _{\hat r\hat \theta } \epsilon _s  + \frac{C}{2}\frac{a}{b}r\sigma 'e^{\left( {C - B} \right)\sigma } \Gamma _{\hat \theta \hat y} \epsilon _s  \nonumber\\
    \delta _\varphi  \psi ^s  &=& \left( {\partial _\varphi  \epsilon _s } \right) + \frac{1}{2}r\sin \theta \dot a\Gamma _{\hat t\hat \varphi } \epsilon _s  - \frac{1}{2}\sin \theta \Gamma _{\hat r\hat \varphi } \epsilon _s  - \frac{1}{2}\cos \theta \Gamma _{\hat \theta \hat \varphi } \epsilon _s  + \frac{C}{2}\frac{a}{b}r\sin \theta \sigma 'e^{\left( {C - B} \right)\sigma } \Gamma _{\hat \varphi \hat y} \epsilon _s  \nonumber\\
    \delta _y \psi ^s  &=& \left( {\partial _y \epsilon _s } \right) + \frac{1}{2}\dot be^{\left( {B - C} \right)\sigma } \Gamma _{\hat t\hat y} \epsilon _s  + \frac{n}{4}j\left( {\partial _y \sigma } \right)\epsilon _s,
\eea
are more problematic. Only the last one, the bulk equation, vanishes naturally if $\epsilon _s  = e^{  \pm \frac{1}{4}j\sigma } \hat \epsilon _s $, where $\hat \epsilon _s$ is a constant spinor. The first four, the brane equations, only vanish if one simply assumes zero spinor components on the brane. Fixing this problem requires the inclusion of the brane's matter content in the action of the theory and deriving a new set of variation equations with extra terms. This is a major endeavor that deserves a separate study. For the purposes of this paper, we consider the above results to be the first order approximation of a more exact supersymmetric solution.

\section{The modified Friedmann equations and final remarks}\label{FriedmannFinal}

Choosing the gauge $b=1$ in the Einstein equations leads to the following results, where we have defined the usual Hubble parameter $H = \left( {{{\dot a} \mathord{\left/ {\vphantom {{\dot a} a}} \right. \kern-\nulldelimiterspace} a}} \right)$:
\bea
    3H^2  &=& 3Ce^{2\left( {C - B} \right)\sigma } F\left( \sigma  \right) - \frac{1}{2}e^{2\left( {C - B} \right)\sigma } \sigma '^2  + \left( {\Lambda  + \rho } \right)e^{2C\sigma }  \nonumber\\
    2\left( {\frac{{\ddot a}}{a}} \right) + H^2  &=& 3Ce^{2\left( {C - B} \right)\sigma } F\left( \sigma  \right) - \frac{1}{2}e^{2\left( {C - B} \right)\sigma } \sigma '^2  + \left( {\Lambda  - P} \right)e^{2C\sigma }  \nonumber\\
    3\left[ {\left( {\frac{{\ddot a}}{a}} \right) + H^2 } \right] &=& \left( {6C^2  + \frac{1}{2}} \right)e^{2\left( {C - B} \right)\sigma } \sigma '^2  + \tilde \Lambda e^{2B\sigma }.\label{modFried}
\eea

Since both sides of the three equations (\ref{modFried}) are dependent on different parameters ($t$ and $y$), then they both must equal a constant. Requiring this, and using $\sigma = \ln h$ along with (\ref{Harmonic}) gives $C= - {1 \mathord{\left/ {\vphantom {1 3}} \right. \kern-\nulldelimiterspace} 3}$ but puts no constraints on the constant $B$. We also choose to define
\bea
    \Lambda  &=& \Lambda _0 e^{ - 2C\sigma }  \nonumber\\
    \rho  &=& \rho _0 e^{ - 2C\sigma }  \nonumber\\
    P &=& P_0 e^{ - 2C\sigma } \nonumber\\
    \tilde\Lambda  &=& \tilde\Lambda _0 e^{ - 2B\sigma }.
\eea

This last implies the following interpretation: What we are studying is a solution that admits multiple branes located at various values of $y=y_I$ ($I = 1, \ldots ,N \in \mathbb{Z}$), such that the harmonic function (\ref{Harmonic}) is modified $Qy + 1 \to \sum\limits_{I = 1}^N {\left( {Q_I {\left| {y - y_I } \right|}  + 1 } \right)} $, or $Q\left[ {\sum\limits_{I = 1}^N {\left| {y - y_I } \right|} } \right] + 1$. As such, the values of $\Lambda$, $\rho$ and $P$ observed from within any of the branes (\emph{e.g.} what we observe in our universe) are dependent on $e^{  2C\sigma }$ evaluated at the specific coordinate of said brane. For simplicity we remove the index 0 from the `bare' values $\Lambda_0$, $\rho_0$, $P_0$ and $\tilde\Lambda _0$. Another way of interpreting this is that the terms $\rho e^{2C\sigma }$, $P e^{2C\sigma }$, $\Lambda e^{2C\sigma }$ and $\tilde \Lambda e^{2B\sigma }$ are evaluated at particular 3-brane slices $y=y_I$, such that $\rho e^{2C\sigma\left(y_I\right) } \rightarrow \rho_I$, etc. Now defining
\be
    \Lambda_{UH}  = \frac{{3Q^2 }}{{2\left( {1 + 3B} \right)^2 }},
\ee
we can write the three modified Friedmann equations (\ref{modFried}) as follows
\bea
    H^2  &=& \frac{1}{3}\left( {\Lambda _{eff}  + \rho } \right)  \label{Friedmann31}\\
    \frac{{\ddot a}}{a} &=& \frac{1}{3}\Lambda _{eff}  - \frac{1}{6}\left( {\rho  + 3P} \right) = H ^2 \quad\quad {\rm where}\quad\quad\Lambda _{eff}  = {\Lambda } - \Lambda_{UH}  \label{Friedmann32}\\
    \tilde \Lambda  &=&  {2\Lambda  _{eff} - 7 \Lambda_{UH} }  + \frac{1}{2}\left({\rho } - {3}P\right).\label{Friedmann33}
\eea

We interpret $\Lambda$ as the cosmological constant related to the value of the quantum mechanical vacuum energy, while $\Lambda _{eff} $ is the observed value of the dark energy density based on the brane-universe's rate of acceleration, where $\Lambda_{UH}$ acts to reduce one into the other, thereby effectively explaining the `vacuum catastrophe'. We get the expected de-Sitter-like spacetime metric on the 3-brane with $a\left( t \right) = e^{Ht} $, as well as the associated equation of state $P=-\rho$. Finally, we note that close inspection of equation (\ref{Friedmann33}) leads to the conclusion that the bulk cosmological constant $\tilde \Lambda$ cannot be zero, but in fact must carry a large \emph{negative} value, implying an anti de-Sitter like behavior in the bulk, which should be accounted for in deeper studies of this result. Furthermore, although the value of the constant $B$ is arbitrary, one can impose the requirement that it gives fields (${\varphi '}$ and ${\chi '}$) that are well-behaved at bulk infinity $y\rightarrow \infty$. This leads to the restriction $B> {{2 \mathord{\left/ {\vphantom {2 3}} \right. \kern-\nulldelimiterspace} 3}}$. The complete solution is then
\bea
    ds_5^2  &=& \left( {Qy + 1} \right)^{\frac{2}{{1 + 3B}}} \left[ { - dt^2  + e^{2 Ht} \left( {dr^2  + r^2 d\Omega ^2_2 } \right)} \right]  + \left( {Qy + 1} \right)^{\frac{{ - 6B}}{{1 + 3B}}} dy^2 \\
    {\rm where}\quad d\Omega ^2_2 &=& d\theta ^2  + \sin ^2 \theta d\phi ^2 \quad {\rm and} \quad  B >   {2 \mathord{\left/ {\vphantom {2 3}} \right. \kern-\nulldelimiterspace} 3}\nonumber\\
    \sigma \left( y \right) &=& \frac{{ - 3}}{{1 + 3B}}\ln \left( {Qy + 1} \right) \\
    \varphi\left( y \right) &=&  \pm \left( {Qy + 1} \right)^{\frac{3}{{1 + 3B}}}  \mp 1 + \varphi _0  \\
    \chi \left( y \right) &=& \frac{1}{2}\chi _0  \left( {1 + j} \right)\left( {Qy + 1} \right)^{\frac{{  3}}{{1 + 3B}}} \nonumber\\
    \bar \chi \left( y \right) &=& \frac{1}{2}\bar\chi _0 \left( {1 - j} \right)\left( {Qy + 1} \right)^{\frac{{  3}}{{1 + 3B}}},
\eea
along with the relationships in the previous set of equations. The constants $\chi_0$ and $\varphi_0$ are the $y=0$ values of the fields. The simplest possible choice of $B$ seems to be the one such that $\sigma \left( y \right) =  - \frac{1}{2}\ln \left( {Qy + 1} \right)$. This is $B =   {5 \mathord{\left/ {\vphantom {5 3}} \right. \kern-\nulldelimiterspace} 3}$, leading to:
\bea
    ds_5^2  &=& \left( {Qy + 1} \right)^{\frac{1}{3}} \left[ { - dt^2  + e^{2 Ht} \left( {dr^2  + r^2 d\Omega ^2_2 } \right)} \right]  + \left( {Qy + 1} \right)^{-\frac{5}{3}} dy^2 \\
    \sigma \left( y \right) &=& {-\frac{1}{2}}\ln \left( {Qy + 1} \right) \\
    \varphi\left( y \right) &=&  \pm \sqrt {Qy + 1}  \mp 1+\varphi_0  \\
    \chi \left( y \right) &=& \frac{1}{2}\chi _0\left( 1+j \right)\sqrt {Qy + 1} \nonumber\\
    \bar\chi \left( y \right) &=& \frac{1}{2}\bar\chi _0\left( 1-j \right)\sqrt {Qy + 1}.
\eea

\pagebreak

\section{Conclusion}

In this work, we have posed a possible resolution to the cosmological constant problem by invoking an extra spatial dimension. We found a solution representing an accelerating brane-universe where the cosmological constant is reduced by the effects of the universal hypermultiplet in the bulk. Of interest is the observation that the solution allows, in fact almost requires, the presence of other parallel brane-universes, even an infinite number of them. The measured values of the cosmological constant, matter density and radiation pressure in each brane are regulated by the dilaton field in the fifth dimension. We also find that a negative cosmological constant is required in the bulk, implying a possible anti de-Sitter like behavior that would be interesting to explore in the future.

As pointed out earlier in detail, the entire result is an approximate first order solution, at least as far as supersymmetry is concerned. One possible scenario of future research is to explore the possibilities of a more exact result, which would involve generalizing the action of the theory in a non-trivial way, as well as exploring the time dependence of the universal hypermultiplet fields; assumed static in this work. Another direction of possible future research is generalizing this model to one with the full set of hypermultiplet fields. Based on previous experience, the presence of non-trivial complex structure moduli of the underlying Calabi-Yau submanifold tends to change the allowed metrics and properties of the solutions (for example, compare solutions found in \cite{1108.3391} and \cite{1208.3488}). We also showed that a second solution is possible based on the choice of a dynamic bulk ($\dot b \ne 0$). We plan to explore this possibility in a separate paper.



\begin{thebibliography}{999}

\bibitem{astro-ph/9805201}
  A.~G.~Riess {\it et al.}  [Supernova Search Team Collaboration],
  ``Observational evidence from supernovae for an accelerating universe and a cosmological constant,''
  Astron.\ J.\  {\bf 116}, 1009 (1998)
  [astro-ph/9805201].

\bibitem{astro-ph/9812133}
  S.~Perlmutter {\it et al.}  [Supernova Cosmology Project Collaboration],
  ``Measurements of Omega and Lambda from 42 high redshift supernovae,''
  Astrophys.\ J.\  {\bf 517}, 565 (1999)
  [astro-ph/9812133].

\bibitem{1203.0307}
  R.~Bousso,
  ``The Cosmological Constant Problem, Dark Energy, and the Landscape of String Theory,''
  arXiv:1203.0307 [astro-ph.CO].

\bibitem{1001.4538}
  E.~Komatsu {\it et al.}  [WMAP Collaboration],
  ``Seven-Year Wilkinson Microwave Anisotropy Probe (WMAP) Observations: Cosmological Interpretation,''
  Astrophys.\ J.\ Suppl.\  {\bf 192}, 18 (2011)
  [arXiv:1001.4538 [astro-ph.CO]].

\bibitem{astro-ph/0208223}
  M.~D.~Maia, E.~M.~Monte and J.~M.~F.~Maia,
  ``The Accelerating universe in brane world cosmology,''
  Phys.\ Lett.\ B {\bf 585}, 11 (2004)
  [astro-ph/0208223].

\bibitem{1006.1850}
  K.~.Saaidi and A.~H.~Mohammadi,
  ``Brane Cosmology for Vacuum and Cosmological Constant Bulk,''
  arXiv:1006.1850 [gr-qc].

\bibitem{hep-th/0009167}
  A.~Falkowski, Z.~Lalak and S.~Pokorski,
  ``Five-dimensional gauged supergravities with universal hypermultiplet and warped brane worlds,''
  Phys.\ Lett.\ B {\bf 509}, 337 (2001)
  [hep-th/0009167].

\bibitem{hep-th/0006215}
  Z.~Kakushadze,
  ``Bulk supersymmetry and brane cosmological constant,''
  Phys.\ Lett.\ B {\bf 489}, 207 (2000)
  [hep-th/0006215].

\bibitem{hep-ph/9910498}
  E.~E.~Flanagan, S.~H.~H.~Tye and I.~Wasserman,
  ``Cosmological expansion in the Randall-Sundrum brane world scenario,''
  Phys.\ Rev.\ D {\bf 62}, 044039 (2000)
  [hep-ph/9910498].

\bibitem{hep-th/9910219}
  P.~Binetruy, C.~Deffayet, U.~Ellwanger and D.~Langlois,
  ``Brane cosmological evolution in a bulk with cosmological constant,''
  Phys.\ Lett.\ B {\bf 477}, 285 (2000)
  [hep-th/9910219].

\bibitem{hep-th/0008192}
  A.~Mennim and R.~A.~Battye,
  ``Cosmological expansion on a dilatonic brane world,''
  Class.\ Quant.\ Grav.\  {\bf 18}, 2171 (2001)
  [hep-th/0008192].

\bibitem{0804.2248}
  D.~S.~Gorbunov and S.~M.~Sibiryakov,
  ``Self-accelerated brane Universe with warped extra dimension,''
  CERN-PH-TH/2008-073 
  arXiv:0804.2248 [hep-th].

\bibitem{hep-th/0611074}
  R.~Koley and S.~Kar,
  ``Braneworlds in six dimensions: New models with bulk scalars,''
  Class.\ Quant.\ Grav.\  {\bf 24}, 79 (2007)
  [hep-th/0611074].

\bibitem{0709.3136}
  R.~u.~H.~Ansari and P.~K.~Suresh,
  ``Bulk scalar field in DGP braneworld cosmology,''
  JCAP {\bf 0709}, 021 (2007)
  [arXiv:0709.3136 [gr-qc]].

\bibitem{hep-th/0308125}
  M.~Bander,
  ``Expanding cosmologies in brane geometries,''
  Phys.\ Rev.\ D {\bf 69}, 043505 (2004)
  [hep-th/0308125].

\bibitem{1108.3391}
  M.~H.~Emam,
  ``Split-complex representation of the universal hypermultiplet,''
  Phys.\ Rev.\ D {\bf 84}, 045016 (2011)
  [arXiv:1108.3391 [hep-th]].

\bibitem{Emam:2010kt}
  M.~H.~Emam,
  ``The Many symmetries of Calabi-Yau compactifications,''
  Class.\ Quant.\ Grav.\  {\bf 27}, 163001 (2010)
  [arXiv:1007.4847 [hep-th]].

\bibitem{Ferrara:1989ik}
  S.~Ferrara and S.~Sabharwal,
  ``Quaternionic Manifolds for Type II Superstring Vacua of Calabi-Yau Spaces,''
  Nucl.\ Phys.\ B {\bf 332}, 317 (1990).

\bibitem{Cecotti:1988qn}
  S.~Cecotti, S.~Ferrara and L.~Girardello,
  ``Geometry of Type II Superstrings and the Moduli of Superconformal Field Theories,''
  Int.\ J.\ Mod.\ Phys.\ A {\bf 4}, 2475 (1989).

\bibitem{0901.3354}
  M.~Vardanyan, R.~Trotta and J.~Silk,
  ``How flat can you get? A model comparison perspective on the curvature of the Universe,''
  Mon.\ Not.\ Roy.\ Astron.\ Soc.\  {\bf 397}, 431 (2009)
  [arXiv:0901.3354 [astro-ph.CO]].

\bibitem{gr-gc/9304026}
  J.~D.~Brown,
  ``Action functionals for relativistic perfect fluids,''
  Class.\ Quant.\ Grav.\  {\bf 10}, 1579 (1993)
  [gr-qc/9304026].

\bibitem{1208.3488}
  M.~H.~Emam,
  ``Zero-branes and the symplectic hypermultiplets,''
  Phys.\ Rev.\ D {\bf 86}, 045016 (2012)
  [arXiv:1208.3488 [hep-th]].

\end{thebibliography}
\end{document}